\documentclass[superscriptaddress,amsmath,amssymb,aps,pra,lengthcheck]{revtex4-1}

\usepackage{graphicx}% Include figure files
\usepackage{dcolumn}% Align table columns on decimal point
\usepackage{bm}% bold math
\usepackage{hyperref}% add hypertext capabilities
\usepackage{amsmath}
\usepackage{amssymb}
\usepackage{dsfont}

\newcommand{\ket}[1] {\vert{#1}\rangle}
\newcommand{\op}[1]{\ensuremath{\Hat{\mathrm{#1}}}}

\def\vec#1{{\boldsymbol{#1}}}  %Vektor
\begin{document}

\title{Quantum phase transition of ultracold bosons in the presence of a non-Abelian synthetic gauge field}

\author{T. Gra\ss}
\affiliation{ICFO-Institut de Ci\`encies
Fot\`oniques, Mediterranean
Technology Park, 08860 Castelldefels (Barcelona), Spain}
\author{K. Saha}
\affiliation {Theoretical Physics Department, Indian Association for
the Cultivation of Science, Kolkata-700032, India.}
\author{K. Sengupta}
\affiliation {Theoretical Physics Department, Indian Association for
the Cultivation of Science, Kolkata-700032, India.}
\author{M. Lewenstein}
\affiliation{ICFO-Institut de Ci\`encies
Fot\`oniques, Mediterranean
Technology Park, 08860 Castelldefels (Barcelona), Spain}
\affiliation{ICREA-Instituci\'o Catalana de Recerca i Estudis Avan\c
cats, Lluis Companys 23, 08010 Barcelona, Spain}

\begin{abstract}

We study the Mott phases and the superfluid-insulator transition of
two-component ultracold bosons on a square optical lattice in the
presence of a non-Abelian synthetic gauge field, which renders a
$SU(2)$ hopping matrix for the bosons. Using a resummed hopping
expansion, we calculate the excitation spectra in the Mott
insulating phases and demonstrate that the
superfluid-insulator phase boundary displays a non-monotonic
dependence on the gauge field strength. We also compute the momentum
distribution of the bosons in the presence of the non-Abelian field
and show that they develop peaks at non-zero momenta as the
superfluid-insulator transition point is approached from the Mott
side. Finally, we study the superfluid phases near the
transition and discuss the induced spatial pattern of the superfluid
density due to the presence of the non-Abelian gauge potential.

\end{abstract}

\date{\today}
\maketitle

\section{Introduction}

Experimental systems involving ultracold bosons in optical lattices
provide us with a unique test bed for studying quantum phase
transitions \cite{bloch1,sachdev} and for mimicking strongly
correlated condensed matter systems \cite{latticereview}. It is
well-known that the simplest of such systems with bosonic ultracold
atoms are well represented by the Bose-Hubbard (BH) model for which
the superfluid-insulator transition has been studied both
theoretically
\cite{fisher,seshadri1,jaksch,dupuis,trefgzer1,trivedi1} and
experimentally \cite{bloch1,ian0}. These ideas have also been
extended to systems of multi-species and higher spin bosons which
provide richer phase diagrams \cite{altman-lukin,isacsson,demler2}.
Such theoretical works have been supported by recent experiments
involving studies of both equilibrium phase transitions and closed
non-equilibrium dynamics in some of these systems
\cite{greiner2,polkovnikov1}.

More recently, there have been several theoretical proposals and
concrete experimental realization of synthetic Abelian vector
potentials for ultracold gases by subjecting them to carefully tuned
Raman lasers (for a review see Ref.\ \onlinecite{dalibard}). These
lasers induce a space or time dependent shift in the energy
dispersion of the atoms at a given momenta and thus emulate the
effect of a synthetic magnetic/electric field
\cite{jakschzoller,juzeli2004,clarkPRL,oktel2,duan,lin,spielmanPRL}.
It has been shown that spatially varying Abelian gauge potentials
acting on a bosonic gas within an optical lattice have a profound
effect on both the superfluid-insulator transition and the
superfluid (SF) state to which the transition takes place
\cite{powellPRL,sengupta,saha1}. In particular, the positions of the
precursor Mott peaks shift to finite momenta and the SF density
develops a spatial ordering pattern which can be detected in
standard experiments. Similar, albeit more complicated, theoretical
proposals have been put forward for realization of synthetic
$SU(2)$ gauge potentials for neutral atoms
\cite{osterloh,fleischhauer,anjaPRA}. However, the effect of
non-Abelian gauge potentials on the superfluid-insulator transition
of ultracold bosons has not been studied so far.

In this paper we consider a two-species boson system described by a
BH model in a square optical lattice and study the effect of a
$SU(2)$ gauge potential on its superfluid-insulator transition. Our
work involves an extension of the resummed hopping expansion in
Ref.\ \onlinecite{ednilson,barry} and is somewhat similar to the
strong coupling expansion technique of Ref.\ \onlinecite{dupuis}.
Here these methods are applied to the case of two-species bosons in
the presence of an $SU(2)$ gauge field. The main results of this
work are the following. First, we show that even for the simplest
non-Abelian gauge field, described by a constant $SU(2)$ gauge
potential, the interplay of inter- and intra-species interactions
and the presence of the external gauge field leads to qualitative
changes in several aspects of the superfluid-insulator transition.
Second, we compute the momentum distribution of the bosons in the
Mott phase near the superfluid-insulator transition and show that
the precursor peaks occur at finite, rather than zero, momenta due
to the presence of the gauge field.  We find a sudden change of the
peak positions when the strength of the gauge field is varied in certain
parameter regimes. This indicates that a slight change in the gauge field strength has strong impact on the
dynamical behavior of the system and might be especially relevant
for its non-equilibrium dynamics. This behavior is somewhat
reminiscent of the QPT of excited states discussed in Ref.\
\onlinecite{esqpt} and of the abrupt sign change of the Hall
conductivity which has recently been found for a system of hardcore
bosons \cite{auerbach}. Third, we demonstrate that the
superfluid-insulator phase boundary displays a non-monotonic
dependence on the strength of the gauge field leading to re-entrant
superfluid-insulator transitions with the variation of the strength
of the gauge field for a fixed hopping strength. Finally, we
construct an effective Landau-Ginzburg theory for the
superfluid-insulator transition and use it to chart out the nature
of the SF phase into which the transition takes place. We
show that for a constant non-Abelian gauge field, the SF density
near the transition does not exhibit any spatial ordering. This
feature is to be contrasted with the case where an Abelian flux
(with half flux quanta per lattice plaquette) is added over the
existing $SU(2)$ potential leading to a spatial pattern in the
SF density.

The organization of the rest of the paper is as follows. We outline
the model describing our system in Section \ref{model}. The Mott
insulating (MI) phase is analyzed in Section \ref{MI}. This is
followed by the analysis of the SF phase in Section \ref{SF}.
Finally, we conclude in Sec.\ \ref{conclusion}.

\section{Model \label{model}}

It is well-known that a $SU(2)$ gauge field can be generated for a
system of two-species ultracold bosons by distinguishing between the
atoms in two different Zeeman levels representing the two 'flavors'
of the non-Abelian theory \cite{osterloh}. In doing so, one
substitutes the standard hopping process by a laser-assisted
tunneling, which may depend on the position and the state of the
atom. One of the crucial features of such a tunneling is its ability
to flip the state of the atom. Taking into account local
interactions between the atoms, the effective Hamiltonian describing
the system can be seen to be the same as that of a two-species BH
model \cite{altman-lukin,isacsson,krutitsky}, with an additional
non-Abelian vector potential in the hopping term providing the
additional inter-species coupling. Defining the number operators
$\op{n}^{\mathrm{a}}_i=\op{{a}^\dagger}_i\op{a}_i$ and
$\op{n}_i^{\mathrm{b}}=\op{b}^\dagger_i \op{b}_i$, where $\op{a}_i$
and $\op{b}_i$ denote the boson annihilation operators of the two
species, the local part of the Hamiltonian reads
\begin{align}
\label{H0}
 \op{H}_0 = \sum_i  \Big[ & \frac{U^{\mathrm{aa}}}{2} \op{n}^{\mathrm{a}}_{i}
(\op{n}^{\mathrm{a}}_{i}-1) +
\frac{U^{\mathrm{bb}}}{2} \op{n}^{\mathrm{b}}_{i} (\op{n}^{\mathrm{b}}_{i}-1)
\\ \nonumber &
+ U^{\mathrm{ab}} \op{n}^{\mathrm{a}}_{i} \op{n}^{\mathrm{b}}_{i} -
\mu^{\mathrm{a}}
\op{n}^{\mathrm{a}}_{i} - \mu^{\mathrm{b}}
\op{n}^{\mathrm{b}}_{i}
\Big],
\end{align}
where $U^{\mathrm{xy}}$ is the strength of interactions between a
pair of particles with flavors ${\rm x}$ and ${\rm y}$ and
$\mu^{\mathrm{x}}$ denotes the chemical potential of species ${\rm
x}$. The Hamiltonian $\op{H}_0$ is easily diagonalized using the Fock state basis:
$\op{H}_0 \ket{n^{\mathrm{a}},n^{\mathrm{b}}} =
E_{n^{\mathrm{a}},n^{\mathrm{b}}}
\ket{n^{\mathrm{a}},n^{\mathrm{b}}}$. It is easy to see that
$\op{H}_0$ allows for ground-state degeneracies which are lifted in
the presence of a hopping term leading to different types of
``magnetic'' orderings in the Mott state\cite{altman-lukin}.

As we wish to focus on the influence of gauge fields on the MI-SF
transition, we shall first consider the parameter regime for which
$\op{H}_0$ has a unique ground-state in the local limit. The
simplest choice in this regard is to assume two independent,
identical systems, i.e. $U^{\mathrm{aa}}=U^{\mathrm{bb}}\equiv U$,
$\mu^{\mathrm{a}}=\mu^{\mathrm{b}}\equiv\mu$, and
$U^{\mathrm{ab}}=0$. With this, $\op{H}_0$ describes a system which
in its ground-state is in both flavors occupied by an equal number
$n$ of particles, with $\mu/U < n < \mu/U +1$ as in the
one-component BH model. This setup will be further investigated in
Sections \ref{CaseA} and \ref{SF}, for the MI and SF phase,
respectively.

In another scenario, investigated in Section \ref{CaseB} and
\ref{SF}, we take into account repulsive interactions between the
components ($U^{\mathrm{ab}} > 0$). In this case, a subspace spanned
by all states $\ket{n^{\mathrm{a}}, n^{\mathrm{b}}}$ with
$n^{\mathrm{a}} + n^{\mathrm{b}} = \tilde n$ forms the degenerate
ground state manifold, where $\tilde n$ is the particle number per
site. For $0 < \mu/U < U^{ab}/U$, one single atom occupies each
site, such that any linear combination $c_1 \ket{1,0} + c_2
\ket{0,1}$ at every site is a ground state of the unperturbed
Hamiltonian $\op{H}_0$. The hopping lifts this degeneracy. As shown
in Ref.\ \onlinecite{isacsson}, for small $\lambda \equiv U^{ab}/U$,
an antiferromagnetic ordering is preferred, while for $\lambda \sim
1$, the system chooses a translational-invariant ferromagnetic
phase. Between these two limits an XY ordering with $c_1 = c_2 =
1/\sqrt{2}$ occurs. In the rest of this work, we shall focus on the
ferromagnetic and the XY phases, where the ground state preserves
translational symmetry.

The kinetic part of the Hamiltonian is given by
\begin{eqnarray}
\label{H1}
 \op{H}_1 = -\sum_{i,j} (\op{a}_i^\dagger, \op{b}_i^\dagger ) J_{ij}
\left(\begin{array}{c}  \op{a}_j \\ \op{b}_j
 \end{array}\right),
\end{eqnarray}
where $J_{ij} = \delta_{<ij>} J \mathrm{e}^{-i (\vec{A}_j \cdot
\vec{r}_j- \vec{A}_i \cdot \vec{r}_i)}$ is a nearest-neighbor hopping  with a
constant strength $J$, and we have chosen $\hbar=c=1$. The phase
factor associated with the hopping is defined by the gauge potential
$\vec{A}_i$, which we choose to be of the following form
\begin{eqnarray}
\label{A}
 \vec{A}_i = (\alpha
\sigma_y, \beta
\sigma_x +2\pi \vec{r}_i \cdot {\hat e}_x \Phi,0),
\end{eqnarray}
where $\sigma_{x,y}$ are the Pauli matrices, $\Phi$ is an Abelian
flux, ${\hat e}_x$ denotes the unit vector along $x$, $\vec{r}_i$ is the
spatial coordinate of site $i$, and $\alpha, \beta$ are parameters
characterizing the non-Abelian vector potential. Although
interesting anisotropy effects can be expected from choosing $\alpha
\neq \beta$ \cite{anjaPRL,fqhe-gap}, in this work we shall consider
$\alpha=\beta$ for simplicity. With this choice, the intra-species
hopping terms ($\op{a}_i^\dagger \op{a}_j$ and $\op{b}_i^\dagger
\op{b}_j$) become proportional to $\cos \alpha$, while the
inter-species hopping terms with non-Abelian vector potential
($\op{a}_i^\dagger \op{b}_j$ and $\op{b}_i^\dagger \op{a}_j$)
become proportional to $\sin \alpha$. For $\alpha=0$ and
$\Phi\neq0$, we thus recover the Hofstadter problem of a constant
magnetic field perpendicular to the two-dimensional (2D) system in
the Landau gauge \cite{hofstadter}. Note that also in the opposite limit, $\Phi=0$
and $\alpha \ne 0$, where the vector potential becomes constant, the
non-Abelian character of the gauge potential, i.e. $[A_i, A_j] \neq
0$, yields a constant but finite gauge field. For
the Abelian flux, we shall focus on $\Phi=p/q$, where $p$ and $q$ are co-prime integers. Most of our
work has been done for $\Phi=0$ or $1/2$; however, the method developed here can be straightforwardly extended to other values of $\Phi$ as shown in Ref.\ \onlinecite{sengupta}.

\section{Mott insulating phase \label{MI}}

\subsection{Hopping expansion \label{expansion}}
Our approach to study the full Hamiltonian $\op{H}=\op{H}_0 +
\op{H}_1$ is based on a resummed hopping expansion as developed in
Refs.\ \onlinecite{ednilson,barry} for the single species BH model.
In this formalism, one considers the hopping term as a perturbation and focuses
on the imaginary-time evolution of the operators: $\op{a}(\tau)=
\mathrm{e}^{\op{H}_0 \tau} \ \op{a} \ \mathrm{e}^{-\op{H}_0
\tau}$. Introducing artificial sources $j_i^{\mathrm{a}}(\tau)$,
$j_i^{\mathrm{b}}(\tau)$:
$\op{H}_1[\{j\}](\tau) = \op{H}_1(\tau) + \sum_i  \Big( \bar
j_i^{\mathrm{a}}(\tau) \op{a}_i(\tau) + \bar j_i^{\mathrm{b}}(\tau)
\op{b}_i(\tau) + \mathrm{h.c.} \Big)$ with $\beta$ the inverse
temperature, and $\{j\}$ denoting the set of all four currents, the free energy
of the system can be written as a functional of the sources:
\begin{eqnarray}
\label{F}
 {\cal F}[\{j\}] = -\frac{1}{\beta} \mathrm{ln Tr} \Big(
\mathrm{e}^{-\beta \op{H}_0} \op{T_\tau} \mathrm{e}^{-\int_0^\beta
\mathrm{d}\tau \ \op{H}_1[\{j\}](\tau)} \Big),
\end{eqnarray}
where $\op{T}_\tau$ indicates imaginary-time ordering. From this it
can be directly seen that the derivatives $\beta \delta {\cal
F}/\delta \bar j^{\mathrm{a(b)}}_i(\tau) = \langle \op{a}_i(\op
b_i)(\tau) \rangle = \Psi^{\mathrm{a(b)}}_i(\tau)$ yield the order
parameter fields $\Psi^{\mathrm{a(b)}}_i(\tau)$ which vanish within
the MI phase. For a description of the Mott physics, it is thus
sufficient to expand ${\cal F}[\{j\}]$ up to second order in the
currents. Furthermore, as quantum fluctuations of the hopping term
scale down with dimension \cite{vollhardt}, in a two-dimensional
system an expansion of ${\cal F}[\{j\}]$ up to first order in the
hopping strength $J$ and a subsequent resummation is expected to
yield qualitatively reasonable results. This is automatically
achieved by performing a Legendre transformation iteratively in the
hopping, which substitutes the source fields $\{j\}$ by the physical
order-parameter field. Carrying out these steps, as detailed in
Refs.\ \onlinecite{ednilson,barry}, we finally obtain the effective
action of the system up to second order in $J/U$:
\begin{align}
\label{action}
 S[\{\Psi\}]^{\mathrm{MI}}= \frac{1}{\beta} \sum_{i,j}
\bigg(\bar \Psi^{\mathrm{a}}_{i}, \bar
\Psi^{\mathrm{b}}_{i} \bigg)
\Bigg[\left(\hat G^0_{ij}\right)^{-1}
 - J_{ij} \Bigg]
\left(\begin{array}{c}
\Psi^{\mathrm{a}}_{j} \\
\Psi^{\mathrm{b}}_{j}
\end{array} \right)
\end{align}
where $\hat G^0_{ij} = \langle \op{T}_\tau\op{O}_i^\dagger(\tau)
\op{O}_j(\tau') \rangle_0$ is the unperturbed two-point function.
Here the operators $\op{O}_i^\dagger$ and $\op{O}_j$ may be of type
$a$ or $b$, and the thermal average is with respect to $\op{H}_0$.
Note that $\hat G^0$ is a function of $\tau-\tau'$; thus $
({\hat G}^{0})^{-1}$ is most easily found in frequency space. We point out
that an alternative way to deriving the same effective action within
a random phase approximation is described in Ref.\
\onlinecite{dupuis} for the BH model, and can straightforwardly be
generalized to systems with an Abelian gauge field \cite{sengupta}. From
Eq.\ (\ref{action}), we find that ${\hat G} = [({\hat
G}^{0})^{-1}-J_{ij}]^{-1}$ is the resummed two-point function which can
be used, for example, to find the boson momentum distribution
\cite{dupuis,sengupta}
\begin{eqnarray}
n_{\vec{k}}= -\lim_{T\to 0} \frac{1}{\beta} \sum_{\omega_{\mathrm{M}}} {\rm
Tr}[{\hat G}(\vec{k},i \omega_{\mathrm{M}})] \label{momdis1}
\end{eqnarray}
These momentum distributions can be observed in time-of-flight (TOF)
measurements, as we shall discuss later.

The excitation spectra of the bosons can be obtained from the poles
of ${\hat G}$ or equivalently by setting up the equation of motion:
$\delta S[\{\Psi\}]^{\mathrm{MI}} / \delta \bar \Psi^{\mathrm{a,b}}
= 0$. In Fourier space, this reads
\begin{align}
 \label{eqm}
 & \sum_{\vec{k}} \Bigg[
\delta_{\vec{k},\vec{k}'} \hat G^0(\omega_{\mathrm{M}})^{-1}  -
J_{\vec{k},\vec{k}'}
\Bigg]
\left( \begin{array}{c}
        \Psi^{\mathrm{a}}_{\vec{k}'}(\omega_\mathrm{M}) \\
            \Psi^{\mathrm{b}}_{\vec{k}'}(\omega_\mathrm{M})
       \end{array} \right) =0,
\end{align}
where $J_{\vec{k},\vec{k}'}$ is the Fourier transform of $J_{ij}$.
After an analytic continuation to real frequencies $i
\omega_{\mathrm{M}} \rightarrow \omega + i \epsilon$, the solutions
of Eq.\ (\ref{eqm}) yield the dispersion relations. In the following
subsections, we apply this general procedure to specific choices of
gauge potentials and parameters $\lambda$ and $\mu$.

\subsection{Independent Species \label{CaseA}}

The simplest non-trivial parameter choice which we shall treat in
this section corresponds to $\lambda=0$ and $0\le \mu/U \le 1$. In
this case, $\op{H}_0$ describes two independent standard BH systems
with unique non-degenerate ground state having one boson of each
species per site. The coupling between them is provided by the
inter-species hopping terms arising from the non-Abelian gauge
potential. For this case, it is clear that the unperturbed Green function has vanishing off-diagonal terms, i.e.
 $G^0_{12} \sim \langle \op{a}^\dagger(0)
\op{b}(\tau)\rangle_0=0$ and $\hat G^0_{21} \sim \langle \op{b}^\dagger(0)
\op{a}(\tau)\rangle_0=0$. From our
symmetric choice of parameters, it also follows that $G^0_{11}=G^0_{22}$ rendering $ {\hat G}^0 \sim
\mathds{1}_{2\mathrm{x}2}$. Further, the site-factorizable nature of
${\hat H}_0$ guarantees that the diagonal elements of $\hat G^0$ are
given by $\delta_{\vec{k},\vec{k}'}$ times a function
$G^0(\omega_{\rm M})$ of a single Matsubara frequency
\begin{align}
\label{G0}
 G^0(\omega_\mathrm{M}) = \sum_{n,m=0}^\infty
\frac{\mathrm{e}^{\beta E_{n,m}}}{{\cal Z}^0} \left(\frac{n+1}{\Delta_{n+1}-i
\omega_{\mathrm{M}}}
-\frac{n}{\Delta_n -i \omega_\mathrm{M}} \right),
\end{align}
where ${\cal Z}^0 = \sum_{n,m=0}^\infty \mathrm{e}^{\beta E_{n,m}}$,
and $\Delta_n = E_{n,m} -E_{n-1,m}$. In the zero-temperature limit,
the Boltzmann sums in Eq.\ (\ref{G0}) reduce to a single term
corresponding to the ground-state occupation numbers, $n=m=1$. In the
following two subsections, we shall compute ${\hat G}$ for the two
simplest choices of gauge potentials corresponding to a constant
non-Abelian gauge field: the one without Abelian flux, $\Phi=0$,
and that with an Abelian flux $\Phi=1/2$.

\subsubsection{$\Phi=0$}

\begin{figure}
\includegraphics[bb=0 0 300 201,width=0.4\textwidth]{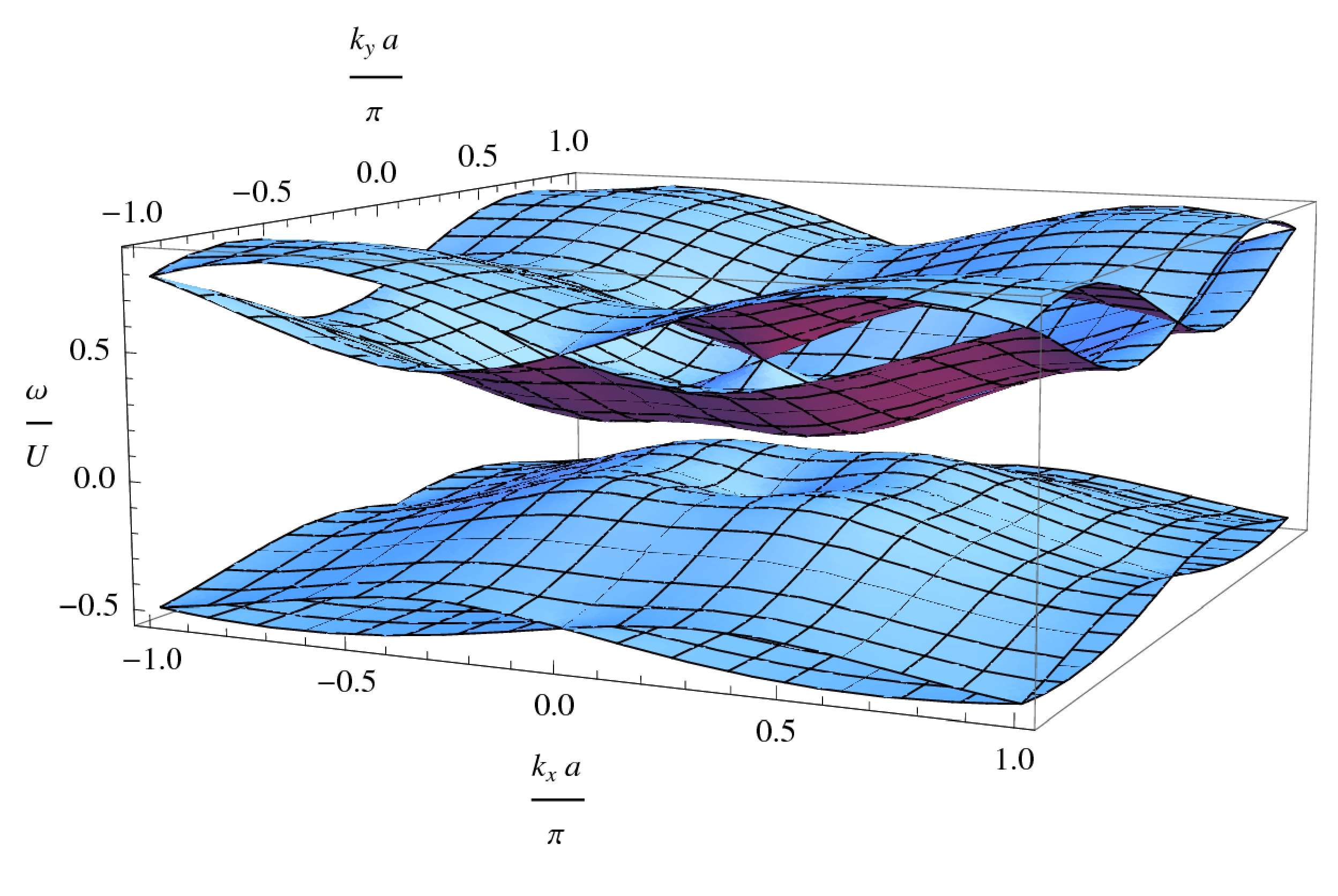}
\caption{\label{spectra} Boson energy dispersion for $\Phi=0$,
$\alpha=1$, $\mu/U=0.41$, and $J/U=0.05$. See text for details.}
\end{figure}

For a constant gauge potential without Abelian flux, $\Phi = 0$, the
hopping matrix is diagonal in momentum space, ${\it i.e.}$,
\begin{align}
\label{Jkk}
 J_{\vec{k},\vec{k'}} = & 2 J \big\{  \cos\alpha [\cos(k_x)  + \cos(k_y)]
\mathds{1}
\\ \nonumber & -\sin\alpha [ \sin(k_x) \sigma_y + \sin(k_y) \sigma_x]
\big\} \delta_{\vec{k},\vec{k}'},
\end{align}
where here and in the rest of the paper, we have set the lattice
spacing $a\equiv 1$. As the unperturbed Green function $\hat G^0$ is
already diagonal, we need to diagonalize only the hopping matrix.
This yields
\begin{align}
E_{\vec{k}}^{\pm} = & 2 J \Big[ \cos(\alpha) \left[ \cos(k_x)+
\cos(k_y) \right] \nonumber \\ & \pm \sin(\alpha) \sqrt{\cos^2(k_x)+
\cos^2(k_y)} \Big]
\end{align}
The energy bands of the system at zero temperature are thus given by
$[G^0(\omega+i\eta)]^{-1}|_{T=0}-E_{\vec k}^{\pm}=0$, where we have Wick rotated
back to real frequency. This yields two quadratic equations
\begin{eqnarray}
\omega^2+\omega(2\mu-U+E_{\vec k}^+)+\mu^2-\mu U+(\mu+U)E_{\vec k}^+&=& 0 \nonumber\\
\omega^2+\omega(2\mu-U+E_{\vec k}^-)+\mu^2-\mu U+(\mu+U)E_{\vec k}^-&=& 0 \nonumber\\
~
\label{bandeqs1}
\end{eqnarray}
leading to four energy bands shown in Fig.\ \ref{spectra}. Two of
these bands occur at $\omega>0$ and the other two at $\omega <0$, so
that they may be interpreted as particle/hole excitation spectra of
the system.  We note that as $\alpha \rightarrow 0$, the particle
and the hole spectra become increasingly similar and ultimately
indistinguishable for $\alpha = 0$ yielding the standard dispersion
of MI bosons with no gauge potential \cite{dupuis}. Furthermore,
the particle and hole excitations have a gap in the Mott phase
which closes as one approaches the superfluid-insulator transition
by increasing $J/U$. Beyond the transition point, which occurs at
$J=J_c$, the solutions of Eq.\ (\ref{bandeqs1}) are complex,
indicating instability of the Mott phase.

\begin{figure}
\includegraphics[width=0.4\textwidth]{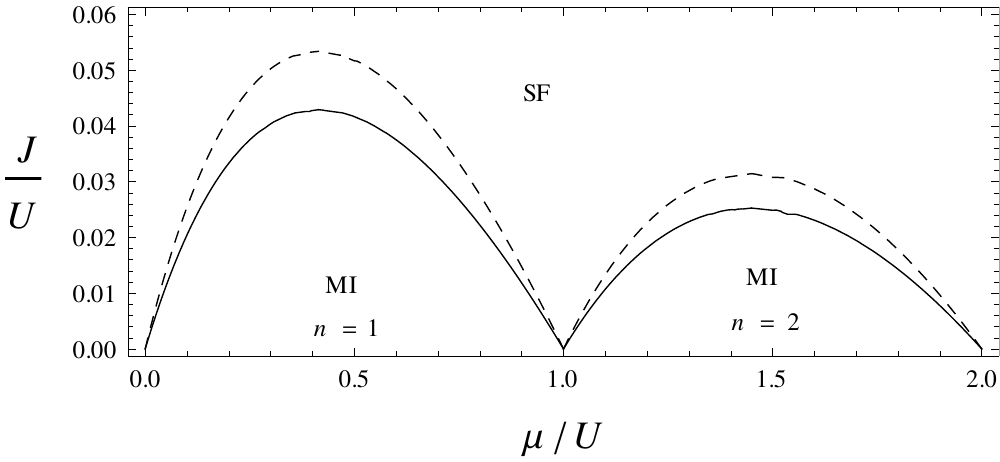}
\includegraphics[width=0.4\textwidth]{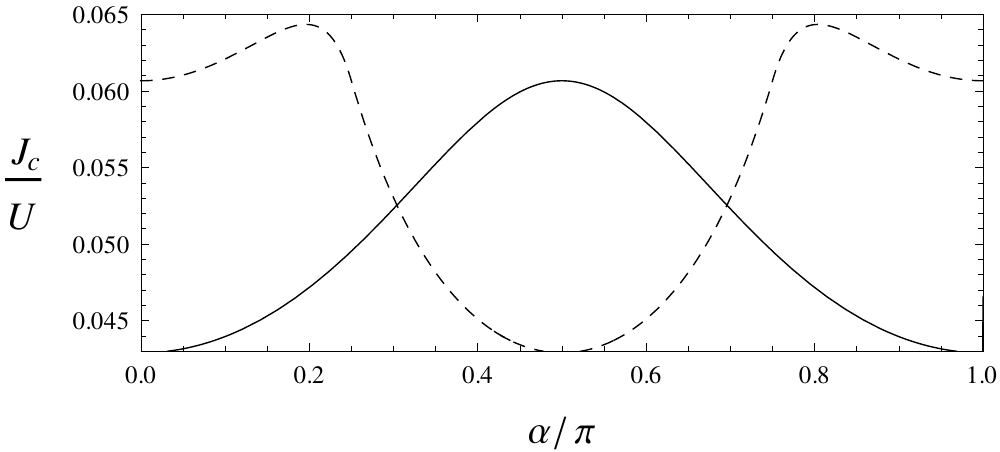}
\caption{\label{boundary} \textbf{Up} The Mott lobes with $\alpha$
dependent heights. The solid (dashed) lines indicate the lobes for
$\alpha=0(1)$. \textbf{Down}: Value of $J_c$ at the tip of the $n=1$
Mott lobe as a function of $\alpha$. The solid (dashed) lines
correspond to $\Phi=0(1/2)$.}
\end{figure}

The superfluid-insulator phase boundary, as obtained by the
procedure described above, is shown in Fig. \ref{boundary}. We find
that the phase boundary has the usual lobe structure. However, the
value of $J_c$ at the tip of the Mott lobe is strongly influenced by
$\alpha$ and can thus be tuned by varying the strength of the gauge
field. This leads to re-entrant superfluid-insulator transitions in
the system by variation of $\alpha$, provided that $J$ is
appropriately fixed at (say) $J=J_c(\alpha=0.2 \pi)$ as can be seen
from Fig.\ \ref{boundary}. We also note that the universality class
of the superfluid-insulator transition has the same properties as in
the standard BH case \cite{fisher}. At the lobe tip, the additional
particle-hole symmetry renders the dynamical critical exponent $z$
of the transition to be unity; at other points, $z=2$.

\begin{figure}
 \includegraphics[width=0.4\textwidth]{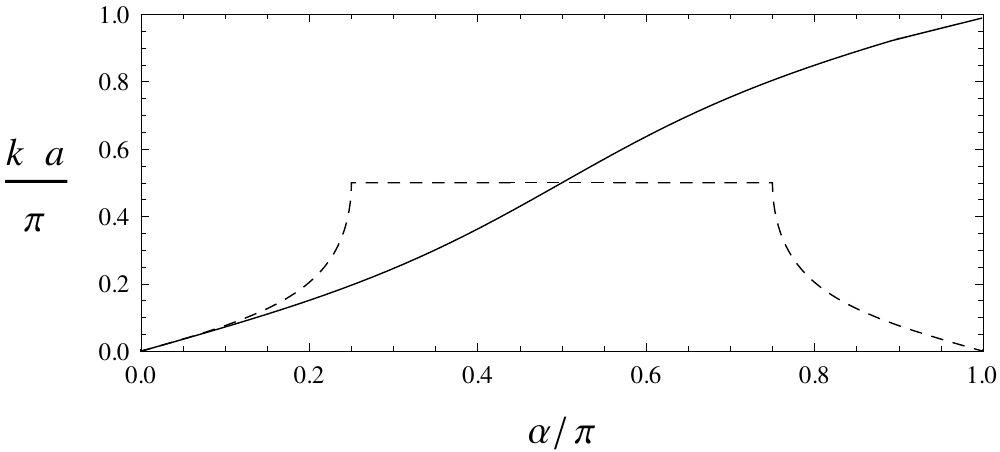}
\caption{\label{kmin} The position of the boson energy minima at
finite wavevector parametrized by $k$ (see text) in systems without
flux (solid line) and with $1/2$ flux (dashed line). In the latter,
a QPT of the excited band occurs at $\alpha=\pi/4$.}
\end{figure}

One of the key difference of the superfluid-insulator transition in
the present system from the normal BH model is that the position of
the minima of the low-energy excitations at and near the critical
point in the first Brillouin zone are at finite momenta and strongly
depend on $\alpha$ (see Fig. \ref{kmin}). Since the position of
these minima correspond to precursor peaks of the bosons near the
critical point \cite{dupuis}, this feature is reflected in the
momentum distribution of the bosons in the Mott phase near the
transition. From the Green function $G$, we compute the momentum
distribution using Eq.\ (\ref{momdis1}). The results of this
calculation are shown in the top panel of Fig.\ \ref{momentum}. We
find that at finite $\alpha$, the precursor peaks of the momentum
distribution at $J \approx 0.97 J_c(\alpha)$ are at finite momenta, reflecting
the fact that the subsequent condensation of the bosons at
$J=J_c(\alpha)$ will occur at non-zero momenta. We note that
such a pattern in the momentum distribution should be easily picked
up in TOF experiments. Since standard TOF experiments will measure
the distribution of both species simultaneously, the presence of
the non-condensing mode will slightly reduce the visibility of the pattern.
However, we expect that one should be able to easily subtract this background in
order to observe the sharp peaks of the condensing mode. The most significant
feature distinguishing the non-Abelian scenario from the known standard one,
is the number of peaks which in general is quadrupled by the
non-Abelian gauge field. We predict this feature to be clearly
observable in TOF experiments.

\subsubsection{$\Phi=1/2$}

Turning to the case with an Abelian flux $\Phi$ in the gauge
potential of Eq.\ (\ref{A}), we find the general structure for the
hopping matrix to be given by:
\begin{align}
J_{\vec{k},\vec{k}'} = A \delta_{\vec{k},\vec{k}'} + B
\delta_{k_y,k_y'}
\delta_{k_x,k_x'+2\pi\Phi} + C \delta_{k_y,k_y'}
\delta_{k_x,k_x'-2\pi\Phi},
\end{align}
making Eq.\ (\ref{eqm}) to have off-diagonal terms connecting momenta
in the magnetic Brillouin zone which differ by $\pm 2 \pi \Phi$.
Here $A$, $B$ and $C$ are functions of momentum which will
be specified later. For a generic $\Phi = p/q$, the periodicity of
the lattice $\Psi_{\vec{k}} = \Psi_{\vec{k}+2\pi n \vec{e}_x}$
ensures that Eq.\ (\ref{eqm}) leads to a set of $q$ closed
equations. To write them down, we introduce the notation
$\vec{\Psi}_{\vec{k}+n\times2\pi\Phi\vec{e}_x}=\vec{\Psi}_{\vec{k},n}$
with $n=0,\cdots,q-1$. In this notation, the equation of motion can
be written in the Harper-like form:
\begin{align}
\label{harper}
M(k_x,n) \vec{\Psi}_{\vec{k},n} + \mathrm{e}^{i a
k_y} N \vec{\Psi}_{\vec{k},n-1} +  \mathrm{e}^{-i a k_y} N^*
\vec{\Psi}_{\vec{k},n+1} =0,
\end{align}
with
\begin{align}
& M(k_x,n) \equiv [G^0(\omega_{\mathrm{M}})]^{-1}
\mathds{1}_{2\mathrm{x}2}
 -2J \\ \nonumber & \times
\Bigg(
\begin{array}{cc}
\cos(a k_x+ 2\pi \Phi n) \cos\alpha & i\sin(a k_x+ 2\pi \Phi n) \sin\alpha \\
-i\sin(a k_x+ 2\pi \Phi n) \sin\alpha  & \cos(a k_x+ 2\pi \Phi n) \cos\alpha \\
\end{array} \Bigg),
\end{align}
and
\begin{align}
 N \equiv -J
\Bigg(
\begin{array}{cc}
\cos\beta & i \sin\beta \\
i \sin\beta  & \cos\beta \\
\end{array}
\Bigg).
\end{align}
As each equation in Eq.\ (\ref{harper}) is a two-component equation,
for a flux of $\Phi=p/q$, we have $2q$ closed equations. Here we
shall focus on $p=1, q=2$ which allows us to find the solutions of
these equations analytically via diagonalization of a $4\times 4$
hopping matrix.

We find that the presence of the magnetic flux $\Phi=1/2$ splits each
band, so that we now have four particle and four hole excitations.
Again the most relevant bands are the particle (hole) excitation at
lowest (highest) frequency. The gap between these excitations closes
at $J=J_c$ leading to a second-order QPT separating the MI and the
SF phases. The lobe structure of the phase boundary (and also the
universality class of the transition) remain unchanged by the flux,
as can be seen from Fig.\ \ref{boundary}. We note however that the
plot of $J_c(\alpha)$ as a function of $\alpha$, shown in the bottom
panel of Fig.\ \ref{boundary}, has a qualitatively different behavior compared to the case without Abelian flux discussed
previously; nonetheless, the system will show similar
re-entrant superfluid-insulator transitions as $\alpha$ is varied
for a fixed $J$.

\begin{figure}
 \includegraphics[bb=0 0 630 327,width=0.5\textwidth]{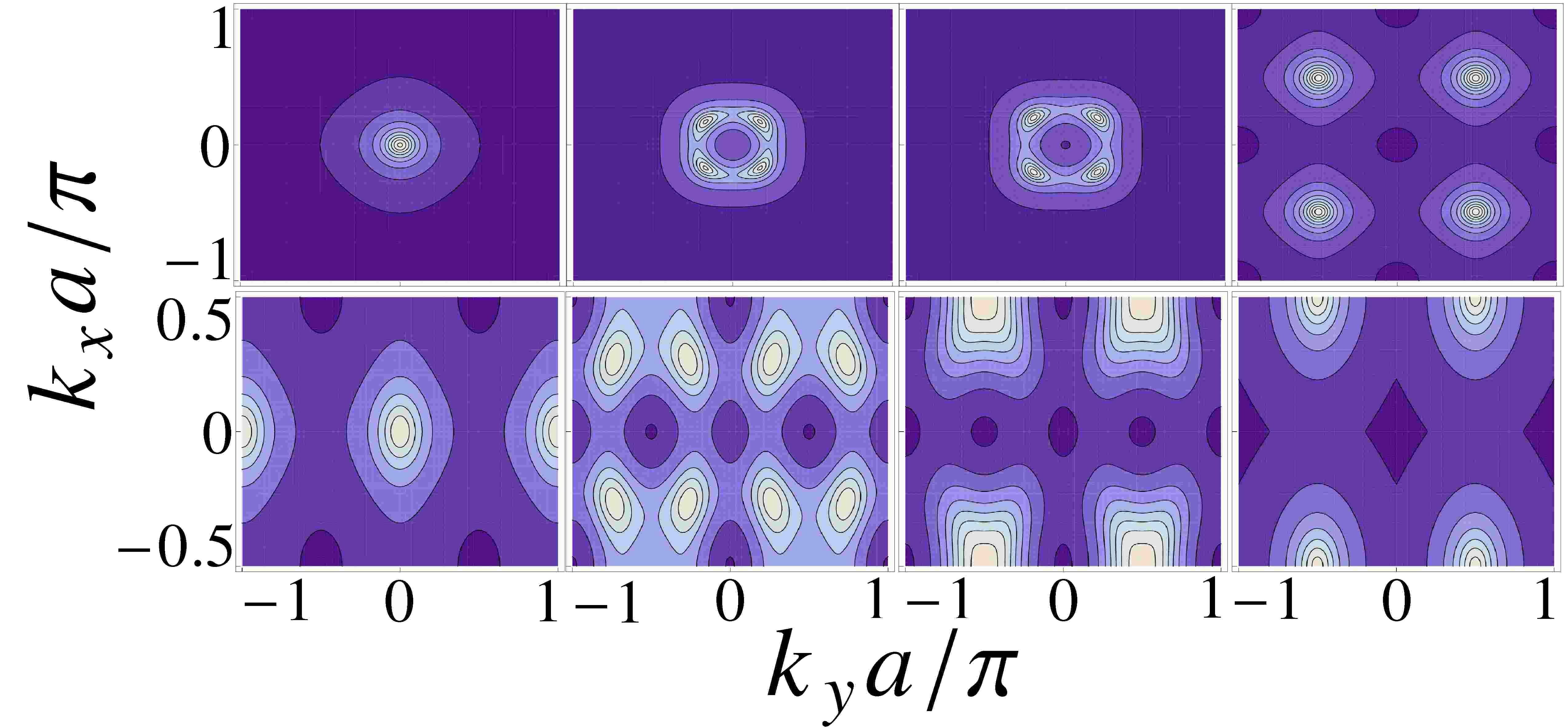}
\caption{\label{momentum} Momentum distributions of the bosons at
$J \approx 0.97 J_c$. Bright regions correspond to high densities. The upper
(lower) rows correspond to $\Phi=0(1/2)$. The non-Abelian field
strengths $\alpha$ are, from left to right in each panel, $0, \,0.7,
\,0.8, \,{\rm and}\, \pi/2$.}
\end{figure}

The most interesting difference between $\Phi=0$ and $\Phi=1/2$
concerns the number and positions of the extrema of the particle and
 hole excitations. We find that within the first magnetic
Brillouin zone ($k_x\in [-\pi/q,\pi/q],\ k_y \in [-\pi,\pi]$) and in
the absence of a non-Abelian field ($\alpha=0$), there are two
extrema at $\vec{k}=(0,0),(0,\pm \pi)$, in agreement with  Ref.\
\onlinecite{sengupta}. Denoting these three points in the Brillouin
zone as $\vec{k}_0$, $\vec{k}_+$, and $\vec{k}_-$, the extrema for
non-zero $\alpha$ can be shown to occur at $\vec{k}_{0} + (\pm k,
\pm k)$, $\vec{k}_+ + (\pm k, -k)$, and $\vec{k}_- - (\pm k, +k)$,
where $k$ as a function of $\alpha$ is plotted in Fig.\ \ref{kmin}.
From this plot, we find that as long as $\alpha <\pi/4$, we have
eight extrema. When $\alpha=\pi/4$, we get $k=\pi/2$ and the extrema
are completely shifted to the zone edges, i.e. again we have only
two extrema per Brillouin zone. For $\pi/4 < \alpha < 3\pi/4$, a
plateau with a single fixed extremum is found.  The derivative
$\mathrm{d}k / \mathrm{d}\alpha$ diverges at $\alpha \rightarrow
\pi/4$ which means that at this critical value of the non-Abelian
field the global minima of the excited bands abruptly change their
position. Thus a slight change in the gauge field strength is
expected to completely modify the dynamical behavior of the system.
This remarkable behavior can be directly observed in the momentum
distribution shown in the bottom panel of Fig.\ \ref{momentum}. In particular, the abrupt change in the
pattern of the momentum distribution when the flux is varied across
$\pi/4$ reflects the sudden change in the position of the
band minima with small change in $\alpha$. This behavior is reminiscent of the
QPT of excited states discussed in Ref.\ \onlinecite{esqpt}.
Also the abrupt sign reversal of the Hall conductivity at half
filling in a system of hard-core bosons subjected to a gauge field
as studied in Ref.\ \onlinecite{auerbach} falls into this category
of phenomena where some control parameter modifies the system's
dynamics in a discontinuous way.

\subsection{XY configuration \label{CaseB}}

In this section, we consider the case where $\lambda >0$ such that
the ground state of the two-species model without the gauge field
correspond to the XY phase discussed in Ref.\ \onlinecite{isacsson}.
where the bosons are in the superposition of both the states.
Consequently, $\hat G^0$ will have non-zero off-diagonal components
and one needs to diagonalize the full matrix $(\hat
G^{0})^{-1}-J_{\vec k}$ in the presence of the non-Abelian flux.

\begin{figure}
\includegraphics[bb=0 0 288 211,width=0.5\textwidth]{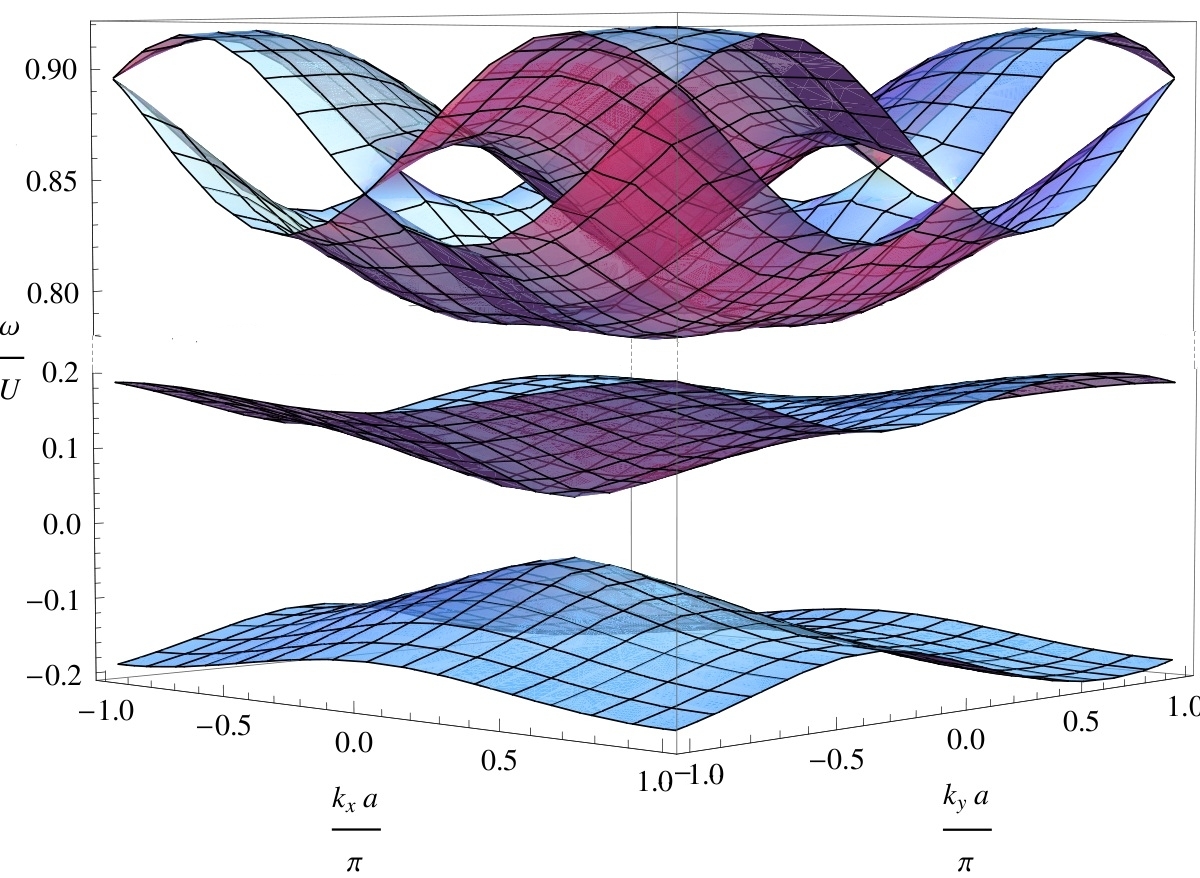}
\caption{\label{spectra2} The excitation spectrum of the bosons in
the XY phase system for $\alpha=1$, $\Phi=0$, $\mu/U=0.15$, $\lambda
= 0.3$, and $J/U=0.021$.}
\end{figure}

We begin by computing the elements of $\hat G^0$ at $T=0$ which are
given by
\begin{eqnarray}
\hat G^0=\left(\begin{matrix}
\langle \op{T}_\tau\op{a}_i^\dagger(\tau)
\op{a}_i(\tau') \rangle_{\mathrm GS}&
\langle \op{T}_\tau\op{a}_i^\dagger(\tau)
\op{b}_i(\tau') \rangle_{\mathrm{GS}}\cr 
\langle \op{T}_\tau\op{b}_i^\dagger(\tau)
\op{a}_i(\tau') \rangle_{\mathrm{GS}}&
\langle \op{T}_\tau\op{b}_i^\dagger(\tau)
\op{b}_i(\tau') \rangle_{\mathrm{GS}}
\end{matrix}\right)\nonumber\\.
\label{gfunc1}
\end{eqnarray}
For the XY ground state
$\ket{\mathrm{GS}}=\frac{1}{\sqrt{2}}(\ket{10}+\ket{01})$,
discussed in Ref.\onlinecite{isacsson}, this matrix reads
\begin{align}
 \hat G^0( i \omega) =& \left[ \frac{1}{2 U \lambda -2 (\mu +i \omega
)}+\frac{1}{U-\mu -i \omega }+\frac{1}{2 \mu +2 i \omega } \right] \mathds{1}
\nonumber \\ &
 + \left[ -\frac{U \lambda }{2 (\mu +i \omega ) (-U \lambda +\mu +i
\omega )} \right] \sigma_x, \label{gfunc2}
\end{align}
where $\mathds{1}$ is the unit matrix and $\sigma_x$ denotes the Pauli matrix.
Inserting Eq.\ (\ref{gfunc2}) into Eq.\ (\ref{eqm}), and considering
the case $\Phi=0$, we find that the presence of the off-diagonal
elements in $\hat G^0$ leads to two independent equations for the band
dispersions given by
\begin{eqnarray}
 M_1+A_k+\sqrt{M_2^2+|B_k|^2+M_2(B_k+B_k^*)}&=& 0 \nonumber\\
 M_1+A_k-\sqrt{M_2^2+|B_k|^2+M_2(B_k+B_k^*)}&=& 0 \label{bandeq2}
\end{eqnarray}
where
\begin{eqnarray}
M_1 &=& \frac{(\omega+\mu-U)(U^2\lambda+U\lambda(\omega+\mu)
-2(\mu+\omega)^2)}{2U^2\lambda-2(\mu+\omega)^2}\nonumber\\
M_2 &=& \frac{U\lambda(\mu+\omega-U)^2}{2U^2\lambda-2(\mu+\omega)^2}\nonumber\\
A_k &=& 2J\cos\alpha(\cos k_x+\cos k_y)\nonumber\\
B_k &=& 2J\sin\alpha(\sin k_y+i\sin k_x) \label{bandeqdef}
\end{eqnarray}

Solving the first of these equations, we find two positive ($E>0$)
and one negative ($E<0$) solutions, while the second equation has solely one positive
solution which is the second highest band. The resulting
bandstructure is shown in Fig.\ \ref{spectra2}. Note that here the
presence of the off-diagonal component of the Green function which
originates from the XY ground state leads to more
particle-like than hole-like excitations. This feature is a
consequence of particle-hole asymmetry originating from $U^{\mathrm{ab}}
\neq U$. Also, the splitting of the two highest particle-like
excitations is a consequence of the non-Abelian nature of the
hopping. This splitting vanishes in the limit $\alpha \rightarrow
0$.

\begin{figure}
\includegraphics[width=0.23\textwidth]{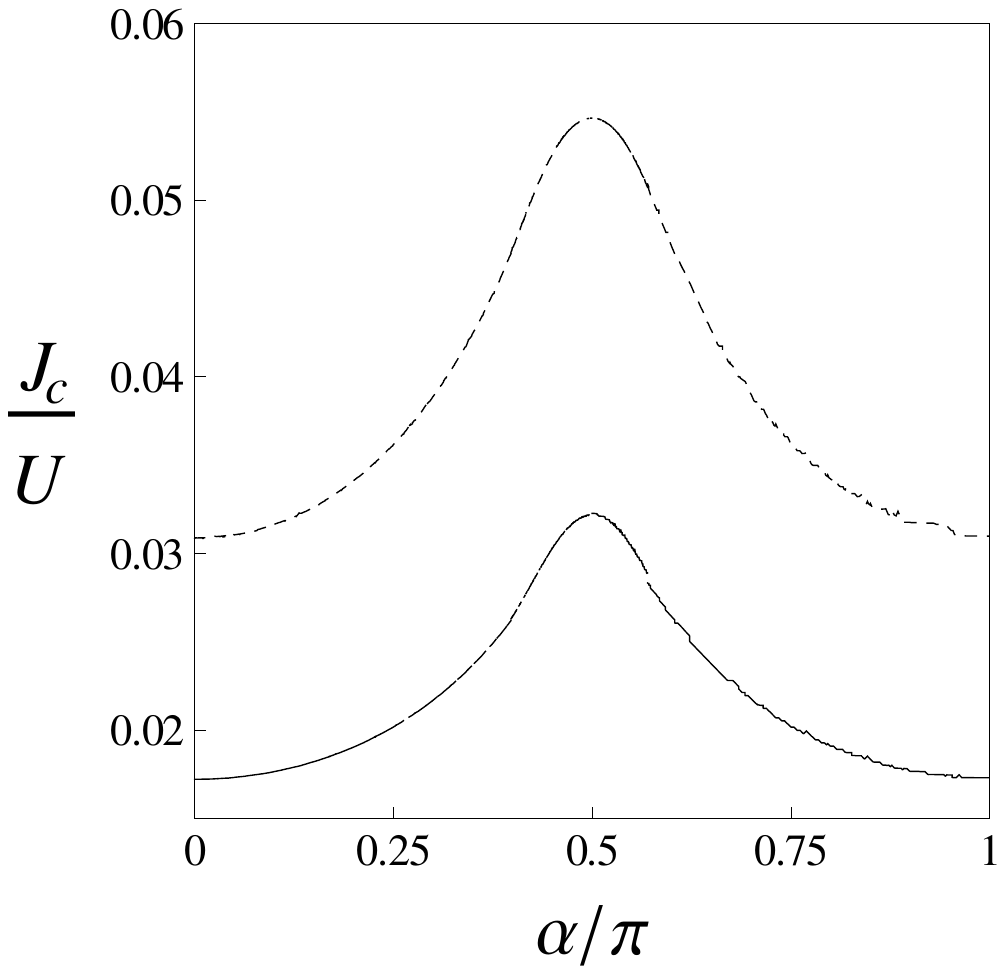} \ \
\includegraphics[width=0.23\textwidth]{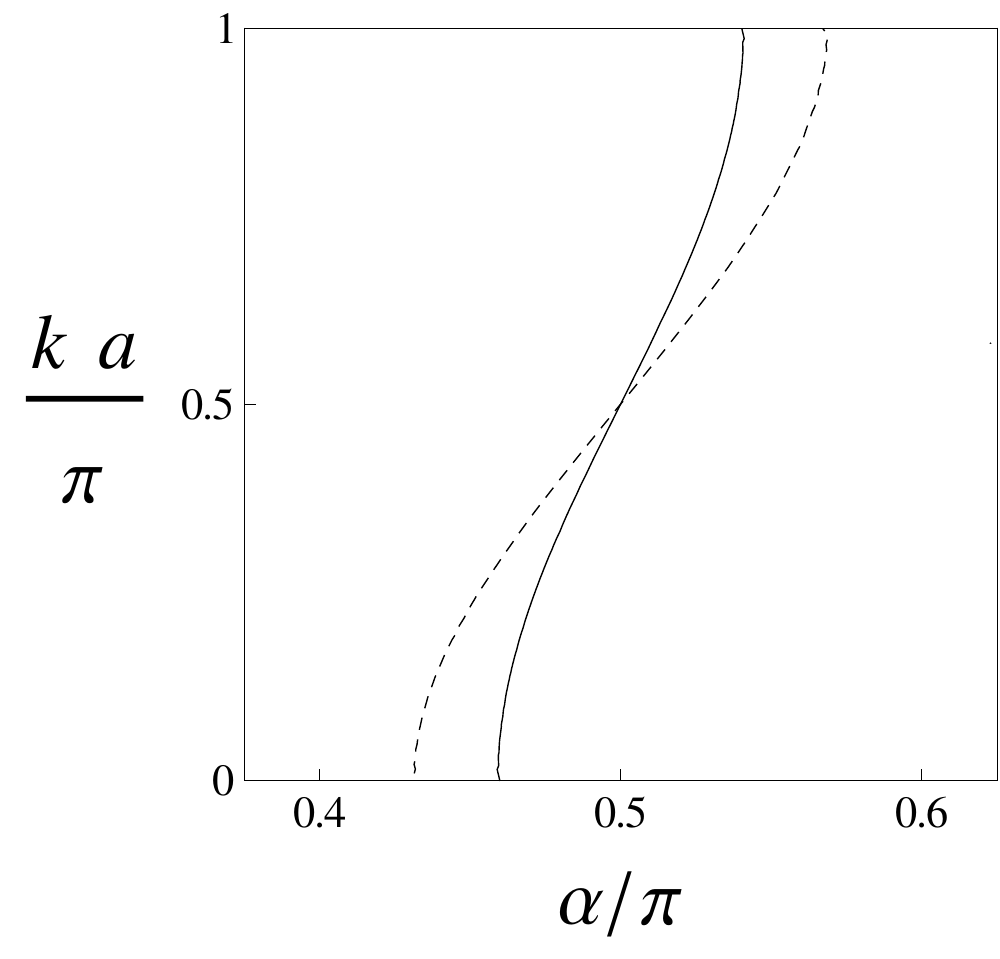}
\caption{\label{kmin2} \textbf{Left}: The critical hopping $J_c$ marking the tip
of the lobe for $\lambda = 0.3$ (solid line) and $\lambda = 0.6$
(dashed line) as a function of $\alpha$. \textbf{Right}: The minima
(maxima) of the lowest (highest) particle (hole) excitation are found
at $k_y a= - \alpha$ and $k=\pm k_x$ as shown in the plot (the solid
line corresponds to $\lambda=0.3$, the dashed line to $\lambda=0.6$). It
is different from 0 or $\pm \pi$ only within a small region around
$\alpha = \pi/2$, in which the single extremum suddenly splits into
two.}
\end{figure}

For the QPT into the SF state these higher modes again do not play
a role. The Mott lobe, on which at least one mode becomes gapless, now extends from
$0<\mu<\lambda U$. The value of $J_c$ marking the height of the lobe depends on both $\alpha$ and $\lambda$, as
illustrated on the left panel of Fig.\
\ref{kmin2}. Furthermore, as in the cases studied before, we find
that the minima of the dispersion occur at finite wavevectors.
However, in contrast to the cases studied before, 
the band spectrum in the XY phase is not symmetric under $k_y \to -k_y$.
This property of the dispersion can be traced back to $G$ since
$B_k$ in Eq.\ (\ref{bandeqdef}) is not invariant under such a
transformation.

\begin{figure}
 \includegraphics[bb=0 0 635 260,width=0.5\textwidth]{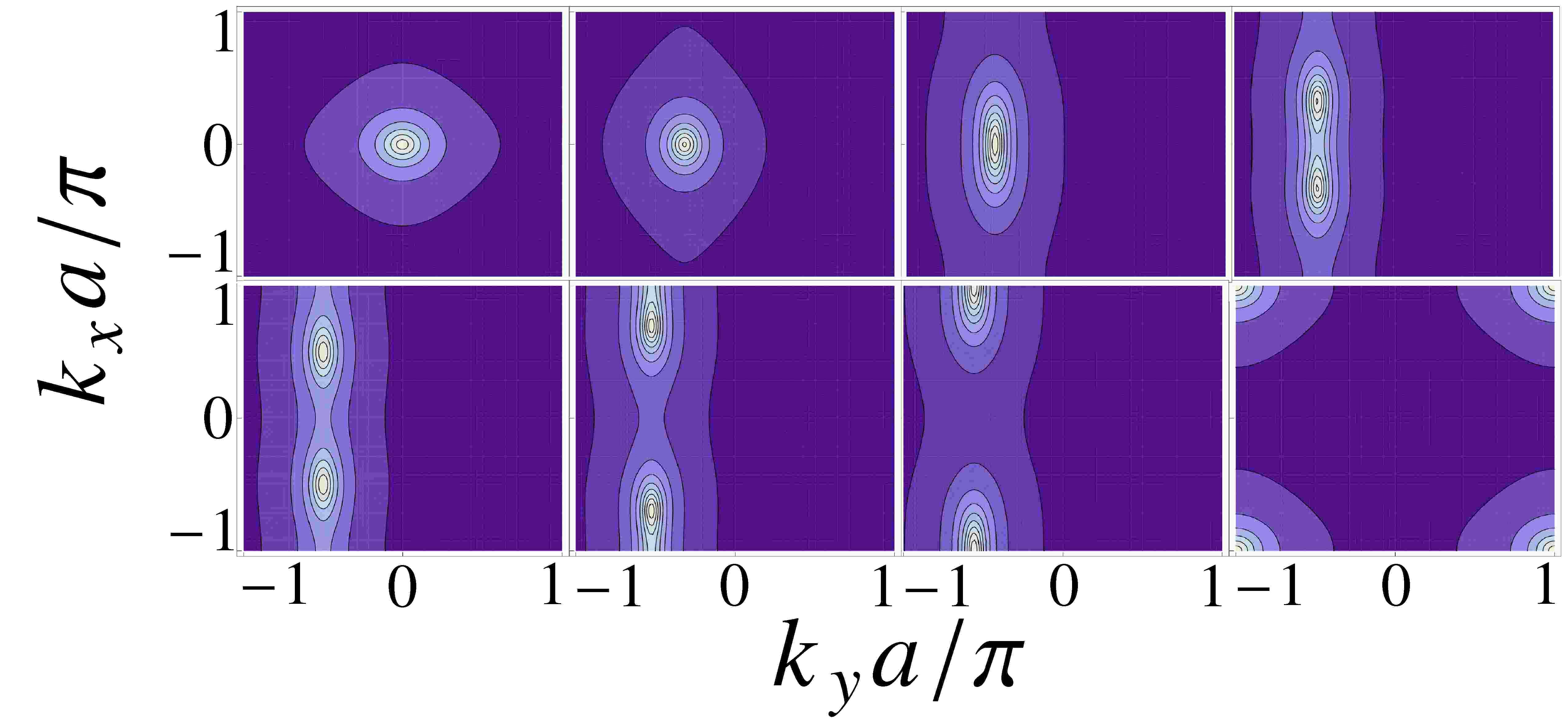}
\caption{\label{momentum2} Momentum distributions at $J \approx 0.97 J_c$ in
XY system for $\lambda = 0.3$. The bright regions correspond to high
densities. The field strengths $\alpha$ are, from left to right,
$0,\, 1, \,1.4, \,1.5, \,\pi/2, \,1.65, \,1.75, \,{\rm and}\, \pi
$.}
\end{figure}

The above-mentioned properties of the dispersion of the particle- and
hole-like excitations are reflected in the nature of the momentum
distribution of the bosons in the Mott phase near the quantum
critical point. We find that for any given $\alpha$, the condensing
modes are located at $k_y =-\alpha$. As shown in Fig.\
\ref{momentum2}, there are no analogous peaks at $k_y = \alpha$
which clearly reflects the breaking of the parity symmetry discussed
above. Also we note that for the range $0\leq \alpha \lesssim 1.5$
($1.65 \lesssim \alpha \leq \pi$), there is a single condensing mode
at $k_x = 0$ ($k_x=\pm \pi$). Around $\alpha = \pi/2$, however, the
condensing mode splits into two at momenta $\vec{k} = (\pm k,
-\alpha)$, where $k$ is plotted on the right panel of Fig.\
\ref{kmin2} as a function of $\alpha$. We note that such a splitting
can modify the dynamical behavior of the system.

\begin{figure}
 \includegraphics[bb=0 0 600 150,width=0.46\textwidth]{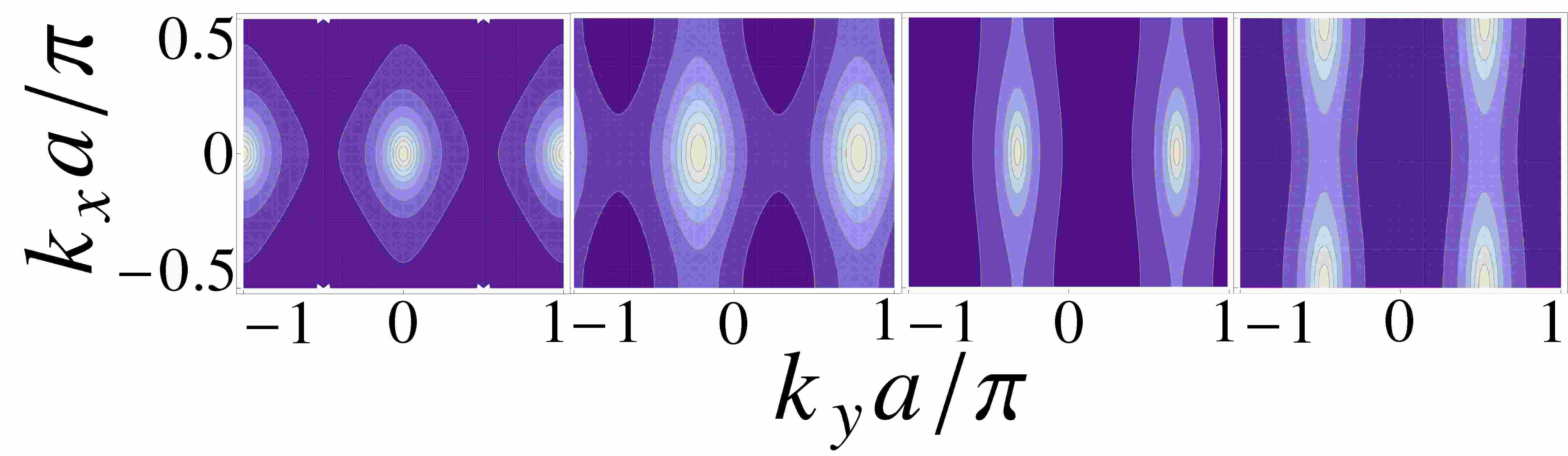}
\caption{\label{momentum3} The momentum distributions of the bosons
at $J \approx 0.97 J_c$ in the XY system with $\Phi=1/2$ for $\lambda = 0.3$ and
field strengths $\alpha$, from left to right, $0,\, 0.7,\, 1, \,{\rm
and}\,1.5$.}
\end{figure}

Finally, we study the influence of a magnetic flux on the XY system,
as done before for $\lambda=0$. As expected we find the Abelian flux
$\Phi=1/2$ to split each band into two, such that the system
exhibits two hole and six particle excitations. As before, for
$\alpha=0$, the presence of this flux provides the band structure with
two extrema at $\vec{k}=(0,0)$ and $(0,\pm \pi)$. The resulting
momentum distribution near condensation is shown in Fig.\
\ref{momentum3}. As $\alpha$ is increased,  the position of the
peaks of the momentum distribution initially shifts along $k_y$.
However, close to $\alpha=\pi/2$, the peaks split along $k_x$
leading to four peaks which finally reach the zone edge
$\vec{k}=(\pm \pi/2, \pm \pi/2)$ at $\alpha=\pi/2$. These features,
as in the case of $\Phi=0$, should be experimentally observable via
standard TOF experiments.

\section{Superfluid Phase \label{SF}}

In this section, we investigate the nature of the SF phase into
which the transition takes place. We note that, as pointed out in
Ref.\ \onlinecite{dupuis,sengupta}, the quadratic action (Eq.\
(\ref{action})) does not capture the physics of the ordered phase but
needs to be supplemented by the quartic term at the mean-field
level. These terms can be directly computed using the methods of
Ref.\ \onlinecite{dupuis,sengupta} within a strong coupling
expansion; however, it is often enough to guess their forms from the
symmetry of the underlying Hamiltonian. In this section, we are
going to take the latter route and chart out the characteristics of
the resulting SF phase.

For determining the order parameters in the SF phase
we need to construct the quartic part of the effective
action in terms of the order parameter fields and minimize it. 
To this end, we first rewrite the quadratic action by
diagonalizing its kernel as shown in Section \ref{MI}. Let us denote
the eigenvalues and eigenmodes of the quadratic action as  $m_{n,\vec{k}}(\omega)$ and
$\Psi_{n,\vec{k}}(\omega)$, respectively. In  the case when the Abelian flux is
$\Phi=p/q$, $n$ varies from $0$ to $2q-1$. Note that here we have
adopted  the convention that for $\Phi=0$, the system remains
with only two eigenmodes, so that we have $n=0,1$. In the
zero-temperature limit we may then write,
\begin{eqnarray}
S^{\mathrm{MI}} = \sum_{n=0}^{2q-1} \sum_{\vec{k}} \int
\mathrm{d}\omega \ m_{n,\vec{k}}(\omega)
|\Psi_{n,\vec{k}}(\omega)|^2.
\end{eqnarray}
Here, the sum over $\vec{k}$ is restricted to the first magnetic
Brillouin zone. The fourth-order term can be written in this basis as
\begin{eqnarray}
S^{(4)} = g/2 \sum_{n=0}^{2q-1} \sum_i \int_0^{\beta}
\mathrm{d}\tau \ | \bar \Psi_{n}(\vec{r}_i,\tau) \cdot
\Psi_{n}(\vec{r}_i,\tau) |^2, \label{s1eq}
\end{eqnarray}
where we have transformed the  $\Psi_{n}$ and $\bar \Psi_{n}$ fields
to real space. Here, $g>0$ is the exact two-particle vertex function
of the bosons in the local limit, which has been computed in Refs.
\cite{dupuis,barry} and $i$ denotes lattice sites.

With this the SF action may be written as
\begin{eqnarray}
S^{\mathrm{SF}} = S^{\mathrm{MI}} + S^{(4)}.
\end{eqnarray}
Now we note that at the onset of superfluidity only one of the
eigenmodes condenses; therefore, it is possible to analyze the SF
phase within a mean-field approximation by ignoring the other modes.
This observation allows us to get rid of the $n$-sum over all bands
in Eq.\ (\ref{s1eq}):
\begin{eqnarray}
S^{(4)} = g/2 \sum_i \int_0^{\beta}
\mathrm{d}\tau \ | \bar \Psi(\vec{r}_i,\tau)
\cdot \Psi(\vec{r}_i,\tau) |^2.
\end{eqnarray}

In the presence of a non-Abelian gauge field with magnetic flux
$\Phi=0$, or flux $\Phi=p/q=1/2$, we have one or more than one
minima of the boson energy spectrum depending upon the non-Abelian
field strength $\alpha$. If the particle/hole modes consist of $s$
degenerate minima, then the corresponding Ginzburg-Landau theory
can be expressed by $s$ low-energy fluctuating two-component
fields (order parameters) $\phi_n(r,t)$ around these
minima \cite{sengupta,saha1}:
\begin{eqnarray}
\Psi(\vec{r}_i,t)=\sum_{n=1}^s\chi_n(\vec{r}_i),
\phi_n(\vec{r}_i,t), \label{Psiasphi}
\end{eqnarray}
where we have Wick rotated to real time. The coefficients
$\chi(\vec{r}_i)$ are the real space eigenfunctions corresponding
to the minimum energy band at
$\vec{k}=(k_x^{\mathrm{min}},k_y^{\mathrm{min}})$ which can be
expressed as:
\begin{eqnarray}
\chi_n(\vec{r}_i)=\sum_{l=0}^{q-1}c_l \mathrm{e}^{(ik_x^{\mathrm{min}}+2
\pi l/q )x_i} \mathrm{e}^{ik_y^{\mathrm{min}}y_i}, \label{rh0}
\end{eqnarray}
Note that the sum in the above expression is restricted to $q$
terms, since the functions $\chi_n(\vec{r}_i)$ describe only the
part of the spatial dependence of $\Psi(\vec{r}_i,t)$ that can be
factored out for each term in the sum in Eq.\ (\ref{Psiasphi}). Here
$c_l$ denotes the components of eigenvectors corresponding to the minimum
energy band at $\vec{k}=(k_x^{\mathrm{min}},k_y^{\mathrm{min}})$.

In general, the quartic part of the Landau-Ginzburg action is
difficult to obtain, since it is restricted only by the invariance
under projective symmetry group (PSG) of the underlying square
lattice \cite{BalentsPSG}. The elements of PSG include in this case
translations along the $x$ and $y$ axes, rotation by $\pi/2$ around
the $z$ axis, and reflections about $x$ and $y$ axes. In our case,
the situation is, however, much simpler, since we know the
microscopic form of the quartic action, Eq.\ (\ref{s1eq}). We may
therefore substitute Eq.\ (\ref{Psiasphi}) into Eq.\ (\ref{s1eq}), and
obtain the explicit form of the quartic action in terms of the order
parameters $\phi_n(\vec{r}_i,t)$. We can then find the saddle point
of the total action with respect to $\phi_n(\vec{r}_i,t)$, and thus
directly calculate the order parameters in the SF phase.

Let us  first consider the case where the number of both
flavors at each site is equal, as discussed in Sec.\ \ref{CaseA}. If the boson
spectrum has one minimum in the magnetic Brillouin zone, then the
corresponding low-energy field can be written as
\begin{eqnarray}
 \psi(\vec{r}_i,t)=\chi_1(\vec{r}_i)\phi_1(\vec{r}_i,t),
\end{eqnarray}
 and thus the SF density reads
\begin{align}
 \rho_s = |\langle \psi\rangle |^2=\left| \sum_{l=0}^{q-1}c_l
\mathrm{e}^{(ik_x^{\mathrm{min}}+2 \pi l/q )x_i}
\mathrm{e}^{ik_y^{\mathrm{min}}y_i}\right|^2|\phi_1|^2.
\label{rh1}
\end{align}
For $\Phi=0$, we can get rid of the sum in Eq.\ (\ref{rh0}) and
$\rho_s$ is simply equal to $|\phi_1|^2$,
which has no modulation along $x$. In contrast, for
$\Phi=p/q=1/2$, we find that $\rho_s$ displays a spatial pattern.

\begin{figure}
 \includegraphics[bb=0 0 360 240,width=.40\textwidth]{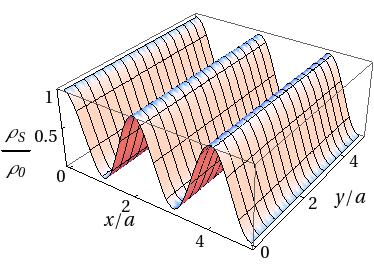} \\
 \includegraphics[bb=0 0 360 274,width=.40\textwidth]{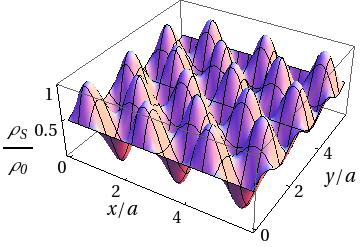}
\caption{a) Plot of SF density $\rho_s/\rho_0$ for
$\Phi=1/2$ and $\alpha=\pi/4$ at $\vec{k}^{\mathrm{min}} \equiv
(\pi/2,\pm \pi/2)$ for even filling. b) Same plot for $\Phi=1/2$
and $\alpha=\pi/2$ at $\vec{k}^{\mathrm{min}}\equiv(\pi/2,\pm\pi/2)$ for XY phase. }
\label{sf}
\end{figure}

Next, we consider the case where there are two minima at
$\vec{k}_1^{\mathrm{min}} = (\pi/2,\pi/2)$ and
$\vec{k}_2^{\mathrm{min}} = (\pi/2,-\pi/2)$ within the magnetic
Brillouin zone for $\alpha=\pi/4$ and $\Phi=1/2$. Note that these
are on the zone edge, such that the minima at the opposite edge are
equivalent. In this case, $\psi(\vec{r}_i,t)=
\chi_1(\vec{r}_i)\phi_1(\vec{r}_i,t)+\chi_2(\vec{r}_i)\phi_2(\vec{r}_i,t)$.
Following the coarse-graining procedure charted out in Ref.\
\onlinecite{sengupta}, we find that the SF ground state corresponds
to the condensation of any one of the low-energy fluctuating fields
$\langle \phi_1 \rangle \ne 0, \langle \phi_2 \rangle=0$ or $\langle
\phi_1\rangle=0, \langle \phi_2\rangle \ne0$. The corresponding plot
for  $\alpha=\pi/4$ and $\Phi=1/2$ in Fig.\ \ref{sf}a, shows a
similar periodic pattern as found for a single minimum for
$\Phi=1/2$. Similar analysis can be done for four minima at $(\pm
\pi/2,\pm \pi/2)$ for $\alpha=\pi/2$ and $\Phi=0$; in this case we
find that only one out the four field condenses; consequently there
is no modulation of SF density. Note that we have restricted
ourselves so far to the minima occurring at the wavevectors
$(\pi/s_1, \pi/s_2)$, where $s_{1,2}=\pm 1$. In principle the
analysis can be extended to the situations when the minima occur at
$(\gamma \pi/2,\delta \pi/2)$ with rational and small $\gamma$ and
$\delta$; however, the analysis becomes technically involved and we
have not attempted that in this work. We do not have general
understanding of implementing the above-mentioned
coarse-graining procedure for irrational $\gamma,\ \delta$.

Finally, we briefly comment on the SF density in the XY phase.
Following the procedure discussed before, we again find a constant
SF density for any non-Abelian gauge field with $\Phi=0$. For
$\alpha=\pi/2$ and $\Phi=1/2$, there are two minima of the spectrum,
and we find that the SF ground state corresponds to the condensation
of both fields around these minima. The corresponding plot for
$\alpha=\pi/2$ is shown in Fig.\ \ref{sf}b.

Thus we generically find that in the presence of a non-Abelian gauge
field, the SF density displays a spatial periodic pattern if there
is a finite flux ($\Phi=1/2$); however, there is no such modulation
without flux, $\Phi=0$. The method that has been discussed here can
be used for any filling fraction $\Phi=p/q$. We expect different
spatial patterns of the SF density for other $\Phi$, and leave the
detailed analysis of it for future study.

\section{Conclusion \label{conclusion}}

In conclusion, we have studied the Mott phases and the
superfluid-insulator transition of strongly-interacting two-species
bosons in the presence of a non-Abelian gauge field. We have shown
that such a system of bosons is expected to display novel patterns of
precursor peaks in the momentum distribution in the Mott phase close
to the superfluid-insulator critical point. We have also
demonstrated the presence of a re-entrant superfluid-insulator
transition as the strength of the non-Abelian field is
varied for a fixed hopping amplitude. Finally, we have found that
the presence of an additional Abelian field with $\Phi=1/2$, leads
to spatial modulation of the superfluid density in the SF phase near
the critical point; however, no such modulation is expected for
$\Phi=0$. We note that the precursor peaks in the momentum
distribution and the presence of the re-entrant superfluid-insulator
transition can be easily detected in standard TOF experiments.
Finally, we point out that we have restricted ourselves to model parameters for which the Mott phase has
translational symmetry; it will be interesting to extend our study
to the case where the Mott state has broken translational symmetry.
We leave this as a subject of future study.

\section*{Acknowledgements}

We acknowledge financial support by ERC Advanced Grant QUAGUATUA, EU Grant STREP
NAMEQUAM, EU IP AQUTE, Spanish MINCIN Grant FIS2008-00784 (TOQATA),
Consolider-Ingenio 2010 QOIT, the Alexander von Humboldt Foundation.
KS thanks DST, India for support under grant SR/S2/CMP-001/2009.

\bibliography{bib}

\end{document}